\documentclass[12pt]{article}
\usepackage{amsfonts}
\def\CC{{\mathbb C}}

          \def\cH{{\cal H}}          
          \def\cK{{\cal K}}           
                    
\def\cP{{\cal P}}                    \def\cR{{\cal R}} 
          \def\cT{{\cal T}}

\def\vR{\check\cR}
\def\qmbox#1{\qquad\mbox{#1}\qquad}
\def\Id{1\hspace{-1mm}{\sf I}}

\def\href#1#2{#2}

\begin{document}

\title{Integrable open boundary conditions for XXC models}
\author{{\bf D. Arnaudon}\thanks{Email address: arnaudon@lapp.in2p3.fr}\\
\\
{\small Laboratoire d'Annecy-Le-Vieux de Physique Th{\'e}orique  
LAPTH\thanks{URA 1436 du CNRS, associ{\'e}e {\`a} l'Universit{\'e} de Savoie},}\\
{\small B.P. 110, F-74941 Annecy-Le-Vieux cedex, France\thanks{Work 
    partially supported by European Community contract TMR 
    FMRX-CT96.0012}}\\
and\\
\\
{\bf Z. Maassarani}\thanks{Email address: zmaassar@phy.ulaval.ca}\\
\\
{\small D{\'e}partement de Physique, Pav. A-Vachon}\\
{\small Universit{\'e} Laval,  Ste Foy, Qc,  
G1K 7P4 Canada}\thanks{Work supported by NSERC 
(Canada) and FCAR (Qu{\'e}bec).} \\}
\date{}
\maketitle

\begin{abstract}
The XXC models are multistate generalizations of the well known
spin-$\frac{1}{2}$  XXZ model. These integrable models share a common
underlying $su(2)$ structure. We derive integrable 
open boundary conditions for the hierarchy of conserved
quantities of the XXC models . Due to  lack 
of crossing unitarity of the $R$-matrix, we develop specific methods 
to prove integrability. The  symmetry of the spectrum is determined.
 
\end{abstract}
\vspace*{1.5cm}
\noindent
\hspace{1cm} September  1998\hfill\\
\hspace*{1cm} LAPTH-695/98, LAVAL-PHY-22/98\hfill\\
%\hspace*{1cm} arch-ive/98mmnnn

\vspace*{1.5cm}
\noindent
%\hspace*{1cm} PACS numbers: ***************\hfill\\
\hspace*{1cm} Key words: Spin-chain, Integrability, 
Open Boundary Conditions\hfill\\

\thispagestyle{empty}

\newpage

\setcounter{page}{1}

\section{Introduction}

Multistate generalizations of the spin-$\frac{1}{2}$  XXZ model were 
studied in \cite{XXC}. These XXC models are not
higher-harmonics generalizations like the spin-$j$ models.
They  have an underlying $su(2)$ structure despite 
exhibiting features related to the $su(n)$ Lie algebras.
The transfer matrix for periodic boundary conditions 
was diagonalized by the algebraic Bethe Ansatz \cite{XXC,ANP}.
Comparing to the spin-$\frac{1}{2}$  XXZ model, the spectrum shows a large 
degeneracy as well as the existence of new eigenvalues in the case of
closed boundary conditions.

In \cite{AARS} the defining Hamiltonians of some multistate extensions
of XXZ models were introduced. For open boundary conditions,
their  spectra were studied with
direct methods,  and without  use of the $\cR$-matrix  approach. 
These models were then also studied in \cite{HakSed}. 
Here, to show integrability,
we  adapt and combine
both approaches, along with a second new  recursion proof.
This is the crux  of this work. 
This specific treatment is required due to an unusual feature
of the XXC $\cR$-matrix: its lack of crossing unitarity.
This property is usually needed to show integrability  for
open boundary conditions.
We start with a short introduction of the general $(\cK,\cR)$ construction
for integrable open boundary conditions. 
A modified construction is introduced 
which  allows to prove the commutation of
the transfer matrices at different spectral parameters. 
The action of the transfer matrix 
on the Hilbert space of the chain is made transparent
using the recursion proof and the algebraic method.
The latter method can  be used to  show the affine symmetry 
of the transfer matrix.

\section{Transfer matrices and local Hamiltonians: the standard
  construction
\label{sect:tr-mat}}

In this section, we recall the standard construction of local
integrable spin chain Hamiltonian from a representation  of the
Yang--Baxter algebra and solutions of the reflection equations. 
We follow the approach and the equations of 
\cite{Cherednik:1984,SKLY,MENE,GouldLinks:1996}. 

The Yang--Baxter algebra reads
\begin{eqnarray}
  && \vR_{i,i+1}(u) \vR_{i+1,i+2}(u+v) \vR_{i,i+1}(v) =   
  \vR_{i+1,i+2}(v) \vR_{i,i+1}(u+v)
  \vR_{i+1,i+2}(u)  \;, \nonumber\\
  && \vR_{i,i+1}(u) \vR_{j,j+1}(v)  = 
  \vR_{j,j+1}(v) \vR_{i,i+1}(u) \qmbox{for} |i-j|\ge 2 \;.
  \label{eq:YBA}
\end{eqnarray}
The $\vR$-matrix with spectral parameter $u$ 
satisfies the inversion relation:
\begin{equation}
  \vR(u)\vR(-u) = \zeta(u) \;.
  \label{eq:inv-rel}
\end{equation}
The matrix $\cR$ is related to the matrix  $\vR$ by $\cR=\cP\vR$,
the operator 
$\cP$ being the permutation map $\cP:x\otimes y\mapsto y\otimes x$.
Usually, $\cR$ satisfies a crossing unitarity relation:
\begin{equation}
  \cR_{12}(u+2\rho)^{t_1} M_1 \cR_{21}(-u)^{t_1} M_1^{-1} = \zeta(u+\rho) \;.
    \label{eq:cross-unit}
\end{equation}
The reflection equations are given by
\begin{equation}
  \cR_{12}(u-v) \cK_1^-(u) \cR_{21}(u+v) \cK_2^-(v) 
  =
  \cK_2^-(v)  \cR_{12}(u+v) \cK_1^-(u) \cR_{21}(u-v) 
  \label{eq:RE-}
\end{equation}
and
\begin{eqnarray}
  && 
  \cR_{12}(-u+v) \cK_1^+(u) M_1^{-1} 
  \cR_{21}(-u-v+2\rho) M_1 \cK_2^+(v) 
  =\nonumber\\
  && \qquad\qquad\qquad\qquad
  M_1 \cK_2^+(v)  \cR_{12}(-u-v+2\rho) 
  M_1^{-1} \cK_1^+(u) \cR_{21}(-u+v) \;.
  \qquad\qquad
  \label{eq:RE+}
\end{eqnarray}

If the monodromy matrix $\cT(u)$ is given by
\begin{equation}
  \cT(u) = \cR_{0L}(u)\cR_{0L-1}(u)\cdots \cR_{01}(u) \;,
  \label{eq:Tu}
\end{equation}
one defines the double row transfer matrix as 
\begin{eqnarray}
  t(u) &=& \mbox{tr}_0\left( \cK_0^+(u) \cT(u) 
  \cK_0^-(u) \cT(-u)^{-1}\right) \\
  &=& \zeta(u)^{-L}\,\mbox{tr}_0 \left(\cK_0^+(u)
  \vR_{L0}(u)\vR_{L-1,L}(u)\cdots\vR_{23}(u)\vR_{12}(u) 
  \right. \nonumber\\  && \ \ \ 
  \left. \times \ \cK_1^-(u) 
  \vR_{12}(u)\vR_{23}(u)\cdots\vR_{L-1,L}(u)\vR_{L0}(u)\right) \; . 
  \label{eq:tu}
\end{eqnarray}
Here $L$ is the number of sites. 

With the relations given above (Yang--Baxter algebra, unitarity and
crossing unitarity for $\vR$; reflection equations for $\vR$ and
$\cK$), one can prove that the transfer matrices for different values
of the spectral parameters commute mutually, i.e.
\begin{equation}
  [t(u),t(v)] = 0 \qquad \forall u,v \;.
  \label{eq:tutv0}
\end{equation}
The normalizations of $\vR$ and $\cK^-$ are not fixed by their
equations. We take for convenience  $\vR(0)=\cK^-(0)=\Id$.
Defining an open chain Hamiltonian as 
\begin{eqnarray}
  H\equiv \left. 
    \frac{dt(u)}{du}
  \right|_{u=0} - 
  \left.
    \frac{d}{du}\mbox{tr}_0\, \cK_0^+(u)
  \right|_{u=0} &=& 
    \mbox{tr}_0 \left( \cK_0^+(0)\right)
  \left(
    2\sum_{j=1}^{L-1} \cH_{j,j+1} +
    \left.
      \frac{d}{du}\cK_1^-(u)
    \right|_{u=0}
  \right)\label{eq:dt/du}\\
  &\phantom{+}& + 2\, \mbox{tr}_0\, ( \cK_0^+(0) \cH_{L0})  \nonumber
\end{eqnarray}
with 
\begin{equation}
  \cH_{j,j+1} = 
  \left.\frac{d}{du}\right|_{u=0} \vR(u)_{j,j+1} 
  \label{eq:Hinteg}
\end{equation}
one is ensured that it will commute with the set of transfer matrices
indexed by the spectral parameter: 
\begin{equation}
  [t(u),H] = 0 \qquad \forall u \;.
  \label{eq:tuH}
\end{equation}
In the precise case presented in the next section, the crossing
unitarity (\ref{eq:cross-unit}) is missing for the proof of the
commutation (\ref{eq:tutv0}). 

\section{The specific model \label{sect:model}}

For the XXC models $\vR$ has   the form \cite{XXC}
\begin{equation}
\vR(u;\gamma)=  P^{(1)}\sin\gamma +P^{(2)} \sin(\gamma-u) + P^{(3)} 
\sin u\label{rcm}
\end{equation}
where $\gamma$ is a `quantum group' parameter and $u$ is 
the spectral parameter.
Let $E^{ab}$ be the $n\times n$ matrix with a one at row $a$ 
and column $b$ and zeros otherwise. 
Let $n$, $n_1$ and $n_2$ be three positive integers such that $n_1+n_2=n$,
and $A_1$, $A_2$ be two disjoint sets whose union is the set of basis 
states of $\CC^n$, with card$(A_1)=n_1$ and card$(A_2)=n_2$. 
The operators $P^{(i)}$ read:
\begin{eqnarray}
P^{(1)}&=&\sum_{a_1\in A_1}\sum_{a_2\in A_2}\left(E^{a_2 a_2}\otimes
E^{a_1 a_1} + E^{a_1 a_1}\otimes E^{a_2 a_2}\right)\\
P^{(2)}&=&\sum_{a_1,b_1\in A_1} E^{a_1 a_1}\otimes
E^{b_1 b_1} + \sum_{a_2,b_2\in A_2} E^{a_2 a_2}\otimes E^{b_2 b_2}\\
P^{(3)}&=&\sum_{a_1\in A_1}\sum_{a_2\in A_2}\left(x_{a_1 a_2} E^{a_2 
a_1}\otimes
E^{a_1 a_2} + x_{a_1 a_2 }^{-1} E^{a_1 a_2}\otimes E^{a_2 a_1}\right)
\end{eqnarray}
The model $(n_1=1,n_2=1)$ corresponds to the spin-$\frac{1}{2}$ XXZ model
whose integrable structure with open boundary conditions has already 
been studied \cite{SKLY}. 

The parameters $x_{a_1 a_2}$ can be removed or, equivalently, set to one in 
the $\cR$ matrix. Define a diagonal operator by
\begin{equation}
F=\prod_{a_1\in A_1} \prod_{a_2\in A_2} (x_{a_1 a_2})^{\frac{1}{2}
(E^{a_2 a_2}\otimes E^{a_1 a_1}-E^{a_1 a_1}\otimes E^{a_2 a_2})}
\end{equation}
The $F$-twist on the $\cR$-matrix \cite{Resh:lmp20,FoeLinRod:1997b},
$\cR^{(x_{a_1 a_2}=1)}(u) = F\,\cR^{(x_{a_1 a_2})}(u)\, F$,
eliminates the twist parameters. 
For {\it open} boundary conditions, this results
in a similarity transformation for  the transfer matrix:
\begin{eqnarray}
  &&t^{(x_{a_1 a_2}=1)}(u) = X\, t^{(x_{a_1 a_2})}(u) \,
  X^{-1}\label{eq:simil}\\ 
  &&X = \prod_{a_1\in A_1} \prod_{a_2\in A_2}  
  \left(x_{a_1 a_2}
  \right)^{\sum_{i<j} E^{a_1 a_1}_i E^{a_2 a_2}_j}
  \label{eq:xto1} 
\end{eqnarray}
where $E^{a_i a_i}_j$ is denotes the $E^{a_i a_i}$ matrix acting on
site $j$, i.e. the occupation number of site $j$ by state $a_i$
(equal to 0 or 1).
A similar transformation was defined in \cite{Dahmen,ACFsl12qs}.

Unless otherwise noted, the twist parameters $x_{a_1 a_2}$ are then
all taken  equal to 1 and the superscript $(x_{a_1 a_2})$ is dropped. 
To simplify the notation we also take, without loss of generality,
$A_1=\{1,...,n_1\}$ and $A_2=\{n_1+1,...,n_1+n_2\}$.
$\vR$ is taken as (\ref{rcm}) divided by $\sin(\gamma-u)$.

\medskip 

For $(n_1,n_2)\neq (1,1)$ 
the general procedure of section \ref{sect:tr-mat} 
fails almost from the beginning
since the   matrix $\vR$  does not satisfy crossing unitarity. The
partially transposed  matrix $\cR^{t_1}$ is  not even  invertible.
This feature was also noticed in \cite{YS} where some Lax pairs 
were found.  
However, we can still define a set of commuting transfer matrices and
an open chain Hamiltonian with local interaction, using information
provided by the underlying spin-$\frac{1}{2}$ XXZ model. 

\section{Transfer matrices and local Hamiltonians: Modified construction}

Let $\tilde \cK^{\pm}$ be 
diagonal solutions of the reflection equations for
the XXZ model for which one  has $M=\Id$ and $\rho=\gamma$ \cite{SKLY}.

We define $\cK^-$ by 
\begin{eqnarray}
  \cK^-_i &=& \tilde \cK^{-}_1 \qmbox{for} 1\le i\le n_1 \;, \\
  \cK^-_i &=& \tilde \cK^{-}_2 \qmbox{for} n_1+1\le i\le n_1+n_2 \;. 
  \label{eq:solK-}
\end{eqnarray}
This matrix $\cK$, together with $\cR$, is solution of the reflection
equation (\ref{eq:RE-}).
Explicitly, 
\begin{eqnarray}
  \cK^-_i &=& \frac{\cos(u+\xi)}{\cos(u-\xi)} 
  \qmbox{for} 1\le i\le n_1 \;,\label{eq:solK-2a} \\
  \cK^-_i &=& 1 \qmbox{for} n_1+1\le i\le n_1+n_2 \;. 
  \label{eq:solK-2b}
\end{eqnarray}
Here $\xi$ is an arbitrary complex parameter. 

The corresponding boundary term then reads 
\begin{equation}
  \frac12 
  \left.
    \frac{d}{du}\cK_1^-(u)
  \right|_{u=0}
  = -(\tan \xi). \sum_{a_1\in A_1}E^{a_1 a_1}_1 
\end{equation}
where the subscript of $E$ indicates that it acts on first site. 

\medskip

The definition of $\cK^+$ is slightly different:
\begin{eqnarray}
  \cK^+_i &=& \frac1{n_1}\tilde \cK^{+}_1 \qmbox{for} 1\le i\le n_1 \;, \\
  \cK^+_i &=& \frac1{n_2}\tilde \cK^{+}_2 \qmbox{for} n_1+1\le i\le
  n_1+n_2 \;. 
  \label{eq:solK+}
\end{eqnarray}
Solutions for $\tilde \cK^+$ being given from those of $\tilde\cK^-$ by 
$\tilde \cK^+ (u)= M \tilde\cK^-(-u+\rho)$, one has 
\begin{eqnarray}
  \cK^+_i &=& \frac1{n_1} \cdot \frac{\cos(-u+\rho+\xi')}{\cos(-u+\rho-\xi')} 
  \qmbox{for} 1\le i\le n_1 \;, \label{eq:solK+2a}\\
  \cK^+_i &=& \frac1{n_2} \qmbox{for} n_1+1\le i\le n_1+n_2 \;. 
  \label{eq:solK+2b}
\end{eqnarray}
Here $\xi'$ is another arbitrary complex parameter. 

The corresponding boundary term 
will then be equal to 
\begin{equation}
  \frac{
    \mbox{tr}_0\, ( \cK_0^+(0) \cH_{L0})
    }
  {
    \left.
      \mbox{tr}_0\, \cK_0^+(u)
    \right|_{u=0} 
    }
  =
  (\tan \xi'). \sum_{a_1\in A_1}E^{a_1 a_1}_L 
\end{equation}
where a term proportional to the identity has been dropped, and
where the subscript of $E$ indicates that it acts on the last site.

The expressions we gave for the integrable boundary terms use formula
(\ref{eq:dt/du}). We will now prove
that it provides an integrable spin chain Hamiltonian.

\medskip

We still consider spectral dependent (double row) transfer matrices as
defined in (\ref{eq:tu}). 
Although our $\cR$ matrix does not satisfy the crossing unitarity
condition, we will prove that these transfer matrices commute for
different values of the spectral parameters. 
We will give two proofs of this argument: one by recursion,
the other using direct algebraic manipulations. 

We want to show that the transfer matrix is  ``order-preserving''
within every set $A_i$. More precisely,
$t(u)|a^{(1)}_1,a^{(2)}_1,a^{(1)}_2,a^{(1)}_3,a^{(2)}_2,\cdots \rangle$, 
$a^{(i)}_j \in A_i$,  is  a linear
combination of states where $a^{(i)}_j$ is to the left of
$a^{(i)}_{j+1}$ for all j's, and fixed $i=1,2$. 
The recursion proof for the action of the transfer matrix uses
the diagonal form of the $\cK^{\pm}$ matrices and the 
specific structure of the $\vR$-matrix.
To this end let momentarily
$e_k$ be  an operator $E^{cd}$ corresponding to the $k^{\rm th}$ `quantum'
{\it i.e.\/} non-auxiliary space 0, and  
define two types of operators for a chain of $i$ sites:
\begin{itemize}
\item $(e_1\cdots e_i)_1$ is order-preserving
\item $(e_1\cdots e_i)^{a_1 a_2}_2$ is such that
$(e_1\cdots e_i)^{a_1 a_2}_2 e_{i+1}^{a_2 a_1}$ is order-preserving
for a chain of $i+1$ sites.
\end{itemize}
Below all non-auxiliary indices are suppressed.
Let
\begin{equation}
\Pi_i(u) = \cR_{0i}(u)\cdots \cR_{01}(u) \cK_0^- (u) \cR_{01}(-u)^{-1}
\cdots \cR_{0i}(-u)^{-1}
\end{equation}
We now show by induction that, for $i \geq 2$,
$\Pi_i(u)$ has the following form:
\begin{eqnarray}
\Pi_i(u) &=& \sum_{c\in A_1 \cup A_2} f_i(u) E^{cc}_0 (e_1\cdots e_i)_1
\label{form}\\
&+& \sum_{a_1\in A_1}\sum_{a_2\in  A_2}\left( 
g_i(u) E^{a_1 a_2}_0 (e_1\cdots e_i)^{a_2 a_1}_2 +
h_i(u) E^{a_2 a_1}_0 (e_1\cdots e_i)^{a_1 a_2}_2 \right) \nonumber
\end{eqnarray}
where all the $u$-dependence is carried by the functions 
$f_i$, $g_i$ and $h_i$. 
These functions may carry an index dependence but
this is irrelevant to the proof.   
An explicit calculation  using
a diagonal but otherwise arbitrary $\cK^-$ matrix, and the explicit 
$\cR$-matrix, shows that $\Pi_2(u)$ has the above form. Assuming
(\ref{form}) holds for a given $i$, one calculates 
$\Pi_{i+1}(u)$ using the explicit $\cR$-matrix and 
find that $\Pi_{i+1}(u)$ also has the form (\ref{form}), for $i+1$.
This concludes the recursion. 
Finally, for $i=L$ and a diagonal $\cK^+$, 
tracing  $\cK^+_0(u) \Pi_L(u)$ over the auxiliary space 0 shows
that only terms of type 1 remain. We have thus proved that 
the transfer matrix is order-preserving. 
If in addition 
$\cK^{\pm}_a=\cK^{\pm}_b$, for all $(a,b) \in A_i\times A_i$, 
then the matrix elements are ``color-blind''.
This means one can calculate the non-vanishing matrix elements,
between a bra and a ket from the Hilbert space of the chain, 
by simply replacing in both states all states of type $A_1$
by one representative $\beta_1$ in this set, and similarly for $A_2$.
This is clear since the functions $f_i(u)$, $g_i(u)$ and $h_i(u)$ are
now color-free. 

We can now conclude that  two transfer matrices $t(u)$ at different spectral 
parameters commute. Taking an arbitrary  matrix element of 
${[t(u),t(v)]}$, introducing the  closure relation of the Hilbert space 
between the two matrices, using the above results reduces the matrix element
to one for the XXZ system. 
The equal weighing of  $\cK^{\pm}$ within the two sets $A_1$ and $A_2$
ensures that the commutation relation is correctly reproduced.
This method is cast below in an algebraic setting.

%% Proof with $J^{a b}_{(r,s)}$

\bigskip

Let $N_1$ (resp. $N_2$) be the number of states of set $A_1$ 
(resp $A_2$) 
in a $L$-site state, i.e. 
\begin{eqnarray}
  N_1 &=& \sum_{j=1}^L \Id_{n_1+n_2}^{\otimes (j-1)} \otimes 
  \left(\matrix{\Id_{n_1} & 0 \cr 0 & 0 \cr}
  \right)
  \otimes \Id_{n_1+n_2}^{\otimes (L-j)} 
  \\
  N_2 &=& \sum_{j=1}^L \Id_{n_1+n_2}^{\otimes (j-1)} \otimes 
  \left(\matrix{0 & 0 \cr 0 & \Id_{n_2} \cr}
  \right)
  \otimes \Id_{n_1+n_2}^{\otimes (L-j)} 
  \label{eq:N12}
\end{eqnarray}

We introduce two families of operators 
$J^{(i) a b}_{(r,s)}$ ($i=1,2$) acting on $L$-site states, 
defined as in \cite{AARS} by:
\begin{itemize}
\item
  $J^{(i) a b }_{(r , s)} (N_i-r) = (N_i-r) J^{(i) a b }_{(r , s)} = 0$,
  i. e. $J^{(i) a b }_{(r , s )}$ vanishes on states with a number of
  states of $A_i$  different from $r$.
\item 
  $J^{(i) a b }_{( r , s )}$ acts as $E^{ab}$ on the
  $s^{\rm{th}}$  site (first site to the left)
  on which the state belongs to $A_i$. (Note that $s \leq  r $ 
  and that $a,b\in\{1,\dots,n_1\}$ if $i=1$ and
  $a,b\in\{n_1+1,\dots,n_1+n_2\}$ if $i=2$.)
\end{itemize}
One can write explicitly
\begin{equation}
  J^{(1) a b}_{(r,s)} = \sum_{{f:\{1,..,r\}\rightarrow\{1,..,L\}\atop
     f \hbox{ \scriptsize increasing}}} E^{ab}_{f(s)}
  \prod_{k=1}^{r} \Id^{(1)}_{f(k)}
  \prod_{j\not\in{\rm Im}(f)} \Id^{(2)}_j
  \label{eq:Jexplicit}
\end{equation}
with $\Id^{(i)}\equiv \sum_{a\in A_i} E^{aa}$, and a similar
definition for $J^{(2) a b}_{(r,s)}$.

We also define $\bar J^{(i) a b}_{(r,s)}$ as acting similarly on $L+1$
sites, including the auxiliary space $0$. 

One can check the following properties: 
\begin{eqnarray}
  && [\bar J^{(i) a b }_{( r , s )} , \vR_{j,j+1} ] = 0 \\
  && [\bar J^{(i) a b }_{( r , s )} , \vR_{L0} ]   = 0 \\
  && [\bar J^{(i) a b }_{( r , s )} , \cK^-_1 ]     = 0 \\
  && [\bar J^{(i) a b }_{( r , s )} , \cK^+_0 ]     = 0 
  \label{eq:comJbar}
\end{eqnarray}
Hence,
\begin{equation}
  [\bar J^{(i) a b }_{( r , s )} ,  
  \cK_0^+(u) \cT(u) \cK_0^-(u) \cT(-u)^{-1} 
  ] = 0 
  \label{eq:comJ2}
\end{equation}

Our aim is now to get rid of the auxiliary space 0 and to prove the
commutation of $J^{(i) a b }_{( r , s )}$ (without bar) with $t(u)$.
One can notice that, on site 0:
\begin{itemize}
\item $\bar J^{(i) a b }_{( r , s )}$   acts only as $E^{ab}$ 
with  $a,b$ both in $A_1$.
\item $\cK_0^+(u) \cT(u) \cK_0^-(u) \cT(-u)^{-1}$ acts only as
    $E^{aa}$, or as $E^{a b}$ with $(a,b)\in A_1\times A_2$ or
    $(a,b)\in A_2\times A_1$.
\end{itemize}

The matrix elements of the commutator (\ref{eq:comJ2}), considered as
an operator acting on site 0, can now be used to prove 
$[J^{(i) a b }_{( r , s )},t(u)]=0$. Let us consider $i=1$ and take first
$c\in A_2$. 
\begin{eqnarray}
  0 &=& 
  \left[\bar J^{(1) a b }_{( r , s )} ,  
  \cK_0^+(u) \cT(u) \cK_0^-(u) \cT(-u)^{-1} 
  \right]  
  \Bigg|_{cc} \\
  &=& 
  \left[\left. \bar J^{(1) a b }_{( r , s )}\right|_{cc} ,  
  \left. 
    \cK_0^+(u) \cT(u) \cK_0^-(u) \cT(-u)^{-1} 
  \right|_{cc}
  \right]  \\
  &=& 
  \left[\Id \otimes J^{(1) a b }_{( r , s )} ,  
  \left. 
    \cK_0^+(u) \cT(u) \cK_0^-(u) \cT(-u)^{-1} 
  \right|_{cc}
  \right]
  \label{eq:comJ3}
\end{eqnarray}
Taking then $c\in A_1$, and $a\neq b \in A_1$
\begin{eqnarray}
  0 &=& 
  \left[\bar J^{(1) a b }_{( r+1 , s+1 )} ,  
  \cK_0^+(u) \cT(u) \cK_0^-(u) \cT(-u)^{-1} 
  \right]  
  \Bigg|_{cc} \\
  &=& 
  \left[\left. \bar J^{(1) a b }_{( r+1 , s+1 )}\right|_{cc} ,  
  \left. 
    \cK_0^+(u) \cT(u) \cK_0^-(u) \cT(-u)^{-1} 
  \right|_{cc}
  \right]  \\
  &=& 
  \left[\Id \otimes J^{(1) a b }_{( r , s )} ,  
  \left. 
    \cK_0^+(u) \cT(u) \cK_0^-(u) \cT(-u)^{-1} 
  \right|_{cc}
  \right]
  \label{eq:comJ4}
\end{eqnarray}
Summing both (\ref{eq:comJ3}) and (\ref{eq:comJ4}) over $c$, on gets 
\begin{equation}
  [J^{(1) a b }_{( r , s )},t(u)]=0
  \label{eq:comJ}
\end{equation}
The same holds for $J^{(2) a b }_{( r , s )}$. 
As in \cite{AARS}, one defines 
\begin{equation}
  J^{(i)}_r(E^{a_1 b_1},E^{a_2 b_2},\dots,E^{a_r b_r}) = 
  \prod_{s=1}^r J^{(i) a_s b_s}_{( r , s )} 
  \label{eq:JEdef}
\end{equation}
with $r=1,...,L$,  and $J^{(i)}_0()$ such that $N_i J^{(i)}_0()=J^{(i)}_0()
N_i=0$. One has 
\begin{equation}
  \Id = \sum_{r=0}^L \sum_{\{a_1,...,a_r\}\in A_i^r}
  J^{(i)}_r(E^{a_1 \beta_i},E^{a_2 \beta_i},\dots,E^{a_r \beta_i})
  J^{(i)}_r(E^{\beta_i a_1},E^{\beta_i a_2},\dots,E^{\beta_i a_r}) 
  \label{eq:somme1}
\end{equation}
$\beta_i$ being a chosen element in $A_i$. 

Let us then consider $t(u)t(v)$,  multiplied on the left by $\Id$
written as in  (\ref{eq:somme1}) for each $i=1,2$. Using the commutation
relations  (\ref{eq:comJ}), one gets
\begin{eqnarray}
  t(u) t(v) &=& \sum_{{r_1}=0}^L \sum_{\{a_1,...,a_{r_1}\}\in A_1^{r_1}}
  \sum_{{r_2}=0}^L \sum_{\{b_1,...,b_{r_2}\}\in A_2^{r_2}} \nonumber\\
  &&
  J^{(1)}_{r_1}(E^{a_1 \beta_1},E^{a_2 \beta_1},\dots,E^{a_{r_1} \beta_1})
  J^{(2)}_{r_2}(E^{b_1 \beta_2},E^{b_2 \beta_2},\dots,E^{b_{r_2} \beta_2})
  \quad t(u)t(v) \nonumber\\
  &&
  \times\,
  J^{(2)}_{r_2}(E^{\beta_2 b_1},E^{\beta_2 b_2},\dots,E^{\beta_2 b_{r_2}}) 
  J^{(1)}_{r_1}(E^{\beta_1 a_1},E^{\beta_1 a_2},\dots,E^{\beta_1 a_{r_1}}) 
  \label{eq:tutv1}
\end{eqnarray}
Under conditions (\ref{eq:solK-2a}--\ref{eq:solK-2b}) and 
(\ref{eq:solK+2a}--\ref{eq:solK+2b}), the 
factor $t(u)t(v)$ in the centre of the formula is then projected
to an equivalent of the 
product $\tilde t(u) \tilde t(v)$ of the XXZ model with boundary
term. Commuting back the 
operators $J$ and using again (\ref{eq:somme1}), one gets 
\begin{equation}
  [t(u),t(v)] = 0 \;.
  \label{eq:tutv}
\end{equation}
Although this result was obtained for $x_{a_1 a_2}=1$, it holds
for arbitrary parameters. This is obvious from the similarity
relation (\ref{eq:simil}). 

\medskip

To conclude this section we have found the action of the
transfer matrix on states involving only one set $A_i$.
We have done so directly and checked the order-preserving
property  for these states, which can serve as  
pseudo-vacua for  a Bethe Ansatz diagonalization. 
For $c_i\in A_1$ one has:
\begin{eqnarray} 
t(u)\, |c_1,c_2,\cdots,c_L\rangle &=&\left[ {\tilde\cK}^+_2(u)\, 
{\tilde \cK}^-_1(u)\,
\frac{\sin^2\gamma}{\sin^2(\gamma-\lambda)}\cdot
\frac{\left(\frac{\sin^2\lambda}{\sin^2(\gamma-\lambda)}\right)^L -1}
{\frac{\sin^2\lambda}{\sin^2(\gamma-\lambda)} -1}\right.\\
& &\left. + {\tilde\cK}^+_1(u)\, {\tilde\cK}^-_1(u)
+{\tilde\cK}^+_2(u)\, \tilde{\cK}^-_2(u)\left( \frac{\sin^2\lambda}
{\sin^2(\gamma-\lambda)}\right)^L \right]
|c_1,c_2,\cdots,c_L\rangle\nonumber
\end{eqnarray}
The equal weights choice made 
for the $\cK^{\pm}_i$ was used but the parameters $x_{a_1 a_2}$ are 
arbitrary.
Thus all states $|c_1,c_2,\cdots,c_L\rangle, \; \forall c_i \in A_1$
share the same eigenvalue. 
For $c_i\in A_2$ the action of $t(u)$ is obtained from the above
with the changes $\tilde\cK_1^{\pm}\longleftrightarrow \tilde\cK_2^{\pm}$.
Again the same degeneracy hold.
Note also that  similar expressions exist for
arbitrary $\cK^{\pm}_i$, with a sum over $A_1$ or $A_2$ in each
of the three contributions.

\section{Symmetries}

A straightforward generalization
of the methods of \cite{AARS}  shows that the  $(n_1,n_2)$-XXC models
have a large affine symmetry.
One first shows  that the  operators $J^{(i)a b}_{(r,s)}$, 
which where seen to commute with the transfer
matrix,  generate  the $L^{\rm th}$ tensorial power of the quantum algebra $U_q(sl(n_1))\otimes U_{q'}(sl(n_2))$,  
for arbitrary $q$ and $q'$. 
This can then  be interpreted 
as an affine symmetry given by:
$U_q(\widehat{sl(n_1)})\otimes U_{q'}(\widehat{sl(n_2)})$.

For arbitrary twist parameters $x_{a_1 a_2}$ the above symmetry is unchanged.
Its generators (and the $J$-operators) are however modified by the 
similarity relation (\ref{eq:simil}). 

{\it Some}  generators of the affine symmetry take a particularly
simple form when all parameters $x_{a_1 a_2}$ are equal to each
other, but not necessarily to one. The operators 
$E^{a_i b_i} = \sum_j E^{a_i b_i}_j$, $(a_i,b_i)\in A_i\times A_i$, $i=1,2$,
then commute with the transfer matrix and generate a  $gl(n_1)\otimes gl(n_2)$ 
sub-algebra of the affine algebra. This small symmetry is  the 
one expected from simple considerations which apply also 
to the closed periodic  XXC chains.

\section{Conclusion}

We have shown that it is possible to construct XXC Hamiltonians 
with integrable open boundary conditions, within the framework
of reflection matrices. A modified construction was necessary 
due to the lack of crossing unitarity of the $\cR$ matrix. 
All the models share the same eigenvalues as the $(n_1=1,n_2=1)$ or XXZ model,
but the  eigenvectors form representations of an affine quantum algebra. 
The methods used here should admit straightforward generalizations
to the  multi-states models of \cite{ANP}.
An analytical or algebraic Bethe Ansatz diagonalization can be carried out
starting from the pseudo-vacua introduced above.
Finally, despite the particular nature of the $\cR$ matrix, it is tempting
to look for a generalization of the crossing symmetry or crossing
unitarity relation which
would permit a proof of integrability along the lines of the standard 
construction. However such a generalization does not seem obvious.

%%\bibliography{bibref,publ}

\begingroup\raggedright\endgroup

\end{document}